**Gas phase condensation of few-layer graphene with rotational stacking faults in an electric-arc**

Soumen Karmakar[1,*], Ashok B. Nawale[2], Niranjan P. Lalla [3], Vasant G. Sathe[3], Vikas L. Mathe[2], Asoka K. Das[4], Sudha V. Bhoraskar[2]

[1]Department of Applied Physics, Birla Institute of Technology (Mesra), Deoghar Campus, Deoghar 814142, India

[2]Department of Physics, University of Pune, Pune 411007, India

[3]UGC-DAE Consortium for Scientific Research, University Campus, Khandwa Road, Indore 432017, India

[4]Laser and Plasma Technology Division, Bhabha Atomic Research Centre, Trombay, Mumbai 400085, India

*Corresponding author.

Tel/Fax: +91-94-72756601. E-mail address: skarmakar@bitmesra.ac.in (S. Karmakar)

**Abstract**

We report the synthesis efficiency of few-layer graphene (FLG) in an external magnetic field modulated DC carbon arc in different non-reactive buffer gases. The effects of buffer gases on the anode erosion rate and the cathode deposit (CD) formation rate have been investigated during the synthesis of FLG. The constituents of the as-synthesized CDs were investigated using transmission electron microscopy, selected area electron diffraction, Raman spectroscopy and X-ray diffraction analysis. A plausible growth mechanism of such FLG is predicted. The results indicate that, under a parametrically optimized condition, an electric-arc of this kind can efficiently generate FLG with rotational stacking faults at a production-rate of few g/min. A guideline for controlling the number of layers of such FLG has also been suggested.

**1. Introduction**

The research on monolayer of carbon made honeycomb two-dimensional lattice structure, scientifically termed as graphene, geared up in lips and bounds since the historical isolation of such layer, from bulk graphite, was reported [1]. Graphene is undoubtedly emerging as the most promising nanomaterial of the 21$^{st}$ century because of its unique blend of unconventional properties [2], which lead to avenues for its exploitation in a wide spectrum of applications [2].

The presence of defects, impurities, grain-boundaries, multiple domains, structural disorders and wrinkles in the graphene sheet can have considerable impacts on its electronic and optical properties. However, with a considerable development in the mono- and bi-layered graphene related research, recent focus has also turned to graphene's few-layer counterparts [2]. The crystallographic stacking of the individual graphene sheets inside the few-layer graphene (FLG) leads to an additional degree of freedom [2]. The distinct lattice symmetries associated with different stacking orders of FLG have been predicted to strongly influence the electronic properties of FLG, including the band structure, interlayer screening, magnetic state, and spin-orbit coupling [2].

Rotational stacking faults in the FLG decouple adjacent graphene sheets so that their band structure is nearly identical to isolated graphene [3]. This is very different from highly oriented pyrolytic graphite (HOPG), where rotational faults are only produced during sample cleaving [4]. Specifically, the Dirac dispersion at the *K*-point is preserved even though the FLG is composed of many graphene sheets. This could explain why magnetotransport [5] and infrared magnetotransmission [6] experiments on *c*-face grown graphene give results very similar to those of an isolated graphene.

It has also been reported that ABA stacked tri-layer FLG is semimetal with an electrically tunable band overlap [7, 8] while its ABC-stacked counterpart is predicted to be semiconductors with an electrically tunable band gap [8].

However, the applications based on these properties is still far from being practically realized because of the unavailability of suitable synthesizing routes with absolute process-product controllability and required production-rate. Another issue of concern, in the synthesis of graphene by conventional methods, involves the use of toxic chemicals and these methods usually result in the generation of hazardous waste and poisonous gases [2]. Therefore, there is a need to explore ways, which can deterministically generate FLG with rotational stacking faults in an eco-friendly manner.

In today's scenario, there is a variety of routes available for the synthesis of FLG [2]. Arc plasma based synthesis of FLG is a relatively new addition to this field and was first brought into limelight by Subrahmanyam *et al* [9] and by the authors [10]. Since then, there has been a deluge in the arc plasma based FLG synthesis [11]; the main advantage of the process being the capability of generating carbon nanostructures at a bulk scale with enhanced crystallinity and lesser defects in an environment friendly manner. Moreover, this process can aid our understanding in molecular dynamics for analyzing the growth mechanism of FLG directly from the carbon vapor-phase.

However, the emphasis of the authors has, so far, been in exploring the possibility of deterministically synthesizing various carbon nanostructures including FLG [10] and carbon nanotubes (CNTs) [12, 13] inside the cathode deposit (CD) by modifying the arc suitably. The current article primarily focuses on exploring how the layers are stacked inside the FLG sheets formed inside a CD synthesized in an external-magnetic-field modulated carbon arc reported elsewhere [10]. Furthermore, the rates of anode erosion and the CD formation, in presence of a number of non-reactive buffer gases, affecting *prima facie* the nucleation and growth of such FLG, have also been examined. The issue of controlling strategy, adopting which, the monitoring of both the staking sequence and the number of layers of such FLG is possible, has also been addressed.

## 2. Experimental details

Initially, the magnetic field modulated arc-plasma reactor, as has been described in detail in our earlier communication [10], was purged with spectroscopic pure buffer gas/ gas mixture to be used for the required operation and evacuated using a rotary vacuum pump till the base vacuum was set to about $10^{-3}$ Pa. Thereafter, feeding the buffer gas/gas mixture the operating chamber pressure was established using a throttle valve. Chilled water was circulated through the reactor jacket at a rate of $3.33\times10^{-4}$ $m^3s^{-1}$ and the temperature of the reactor walls along with the cathode holder was maintained at 298K throughout the experiment. Synthesis of FLG was carried out igniting the arc in between two spectroscopic pure cylindrical glassy graphite rods with the density of $2.2\times10^3$kg $m^{-3}$. The arc current and voltage were kept constant during all the synthesis runs using a current regulated power supply and manual translation of the anode aided by a rack and pinion arrangement. The operating parameters of the six different synthesis runs are tabulated in Table 1.

Table 1. Experimental parameters during the synthesis of FLG.

| | |
|---|---|
| DC arc current | 180±5A |
| DC arc voltage | 21±2V |
| Operational chamber pressure | 3.75±0.075 Pa |
| Initial gas temperature | 298K |
| Final average gas temperature | 332K |
| Separation between magnetic rings (ΔZ) [10] | 60mm |
| Diameter of the anode | 13mm |
| Diameter of the cathode | 30mm |
| Buffer gases (in molar fractions) | 100% Ar, 100% He, 100% $N_2$, 50%Ar+50%He, 50%Ar+50% $N_2$, 50% $N_2$+50%He |
| Run time | ~3 minutes |

The operating pressure, electrode-dimensions and the magnetic field configuration were exactly the same as those which were found to be most suitable for the production of FLG in this reactor, as evident from our earlier report [10]. However, the current and voltage values were slightly modified during the present experiment, in order to maintain stable arc for all the six buffer gases used. However, the input power remained nearly the same as that of our previous work [10].

After completion of each synthesis run, the reactor was allowed to cool down to the ambient and then the throttle valve was opened to increase the chamber pressure to the atmospheric level. The columnar deposits were dismantled from the cathode surface and the soft cares [10] of these deposits were then isolated. These cores were then investigated further in their totality. The weight measurements as reported in the following section were carried out using a digital balance having a least count of 0.1 mg. The CDs were mechanically homogenized by a pestle and mortar. The products were then investigated by Raman spectroscopy and transmission electron microscopy (TEM) exactly in a similar way reported elsewhere [12]. In addition, the crystal structures of the as-synthesized samples were examined by the X-ray diffraction (XRD) analysis.

The TEM analysis was carried out using a 200 kV Tecnai $G^2$ 20 microscope, equipped with a $LaB_6$ filament and a CCD camera. Raman spectra were recorded in backscattering geometry at room temperature with the help of a Jobin Yvon Labram HR800 spectrometer, in which a He–Ne laser ($\lambda$ = 632.81 nm) was used for the Raman excitations. The integration time during the recording of the Raman spectra was 100s. Before testing the as-synthesized samples, the Raman spectrometer was preciously calibrated with the help of a spectroscopic-pure silicon sample. The diameter of the laser spot was about 1μm. SEM and XRD measurements were carried out using Jeol JSM-6360A microscope and Bruker AXS D8 Advance diffractometer respectively.

## 3. Results and discussion

During the process of arcing, the anode got eroded and a part of the eroded material was found deposited on the plasma facing surface of the cathode. The CD was of columnar shape with the typical characteristics reported elsewhere [10]. Quite interestingly, the diameter of all the CDs was found to be ~20±1mm and was unaffected by the specific buffer gas used during a synthesis run. It is obvious from this finding that the diameter of the CD is primarily decided by the dimensions of the electrodes, arc-current and 'strength and orientation' of the magnetic field used, as was demonstrated in our earlier communication [10].

*3.1 Mass measurements*

Masses of the anode and the CD were estimated after each synthesis run and the rates of anode erosion and the CD deposition were calculated using the corresponding run times. Various mass measurements resulted into three noteworthy results, which are summarized in Figs. (1-3). The error-bars, in these figures, represent the statistical fluctuations encountered during the repeated mass-measurements.

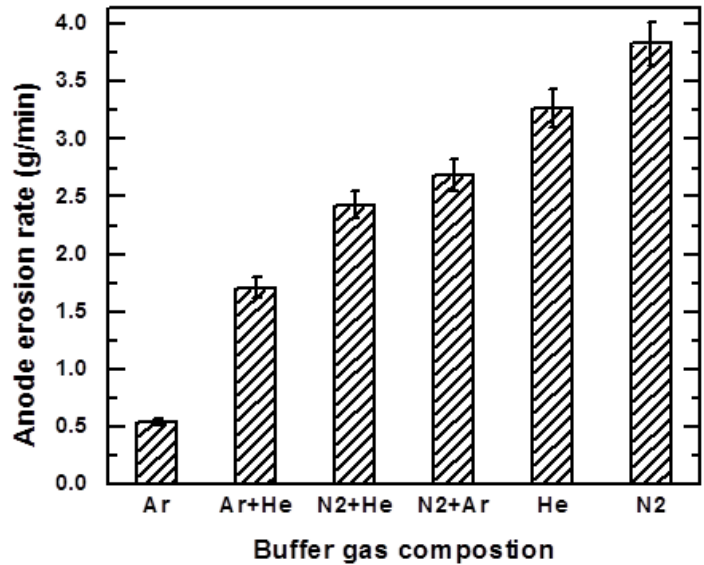

Fig.1. Anode erosion rate for the used buffer gases.

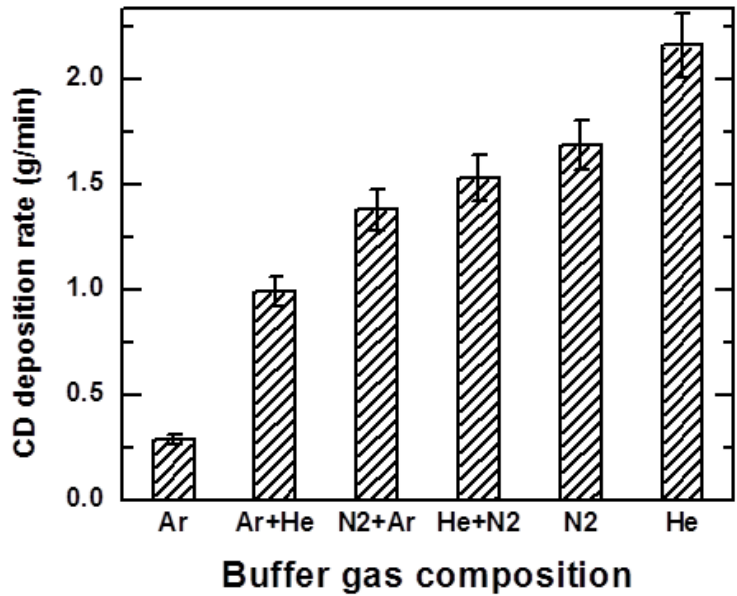

Fig.2. Formation rate of CD for the used buffer gases.

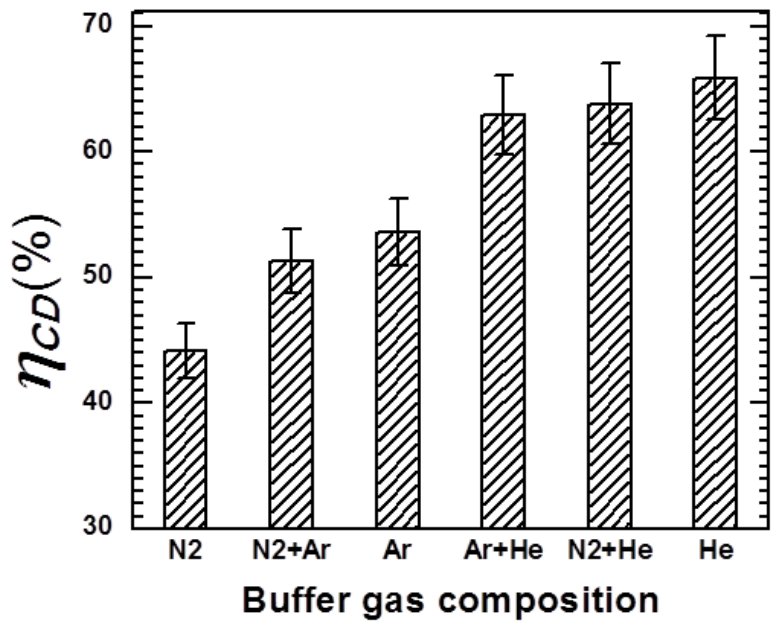

Fig.3. Percentage efficiency ($\eta_{CD}$) of formation of CD for the used buffer gases.

The parameters in Figs.(1-3) have been plotted in the ascending order for the sake of clarity.

It is seen from these figures that the rates of anode-erosion and CD-formation and the conversion efficiency $\eta_{CD}$ [12] of eroded anode material into CD are distinctly affected by the used buffer gases, the thermo-physical properties of which are remarkably different. It was found that the anode erosion rate and CD deposition rate are lowest in case of Ar, while the highest values correspond either to He or $N_2$, the intermediate values being attributed to the mixed buffer gases.

It is obvious that the enthalpy of the plasma state near the anode plays a very important role in deciding the anode erosion rate. It is also known [14] that presence of He in the carbon plasma remarkably enhances the enthalpy content of the plasma; whereas, substantial reduction in enthalpy is observed in case of C-Ar plasma. On the other hand, the CD deposition rate and $\eta_{CD}$ will depend mostly on thermal conductivity and diffusion coefficients of the corresponding buffer gases, which in turn, will depend on the exact temperature profile of the arc. However, it is to be noted that in the near-electrode regions, the plasma is no longer in local thermodynamic equilibrium and the exact proportions of carbon and the buffer gas molecules, as well as the temperature distribution, are pretty difficult to estimate. In view of this, exact reasoning, behind the observed facts revealed through Figs. (1-3), appears to be quite complicated and is beyond the scope of the present communication.

*3.2 TEM analysis*

Figs. (4-9) show a few typical TEM micrographs recorded for the as-synthesized samples. It is evident from these micrographs that all the as-synthesized samples contain three different co-existing nano-dimensional species; namely, tubes, particles or onions and flakes. However, flake like structures appear to be the most abundant species in all the six samples. In order to verify this, the as-synthesized samples were thoroughly examined, in batches, by TEM. Distinct visual differentiation, in terms of the relative content of the flakes, appears to be quite difficult while comparing the six different products under investigation. The fakes are of variable

thickness and in-plane-dimension. On many occasions they are found to scroll to give rise to tubular structures (Fig. 4(b)) as was seen in our earlier study [10].

However, detailed structures of these flakes can be understood analyzing the corresponding SAED patterns as shown in Figs. 4-9(d). The SAED patterns confirm the presence of honeycomb lattice structures of carbon and are marked by the *rare* presence of (*002*) diffraction spots (Fig. 4-9(d)), indicating lesser abundance of closed graphitic structures *viz.* particles and CNTs. This is because, if pristine and flat FLG is seen under TEM, one will, usually, not observe the (*002*) spots. The presence of (*002*) spots or discontinues (*002*) circular ring, as observed in some of the SAED patterns (Fig. 4(d), inset of Fig. 6(c)), is mostly due to the presence of closed graphitic structures like onions and CNTs with the FLG sheets. For $N_2$ and Ar samples the closed graphitic structures appear to dominate along with the FLG sheets. In case of the sample produced in He (Fig. 5(b)), it appears that the FLG sheets are randomly shrunk (like randomly oriented cotton carpet spread on a floor). The FLG sheets made using mixed gases appear to be reasonably flat and free from other graphitic contaminations. This is evident by the near absence of (*002*) reflections from the diffraction patterns corresponding to the mixed buffer gases.

It is to be noted that the presence of FLG can be realized by the occurrence of intersecting elliptical diffraction rings [10], which appear when the electron beam direction, during TEM, is not exactly perpendicular to the FLG sheet. If it happens to be perpendicular to the basal plane of FLG, the rings will be exact-circular as in the case of diffraction rings recorded for the samples prepared in mixed buffer-gas-ambience. The indexing of these rings is illustrated in Fig.7 (d). The presence of FLG can be confirmed by tilting the sample. Under tilted condition the rigs elongate into elliptical rings (with perfect-circular (*000*) spot) having the same indexing as those of the corresponding circular ones. In case of the FLG flakes synthesized in the single gas ambiance, the diffraction patterns (insets of Figs.4(b), 5(a), 6(c)), for the tilted position of the samples, are intersecting ovals and the interpretation of this kind of pattern is available elsewhere [10].

The SAED patterns, corresponding to the graphitic crystals grown in presence of 100% Ar, 100% He and 100% $N_2$ mostly consist of segmented concentric circles or ellipses with radii or the semi minor axes in accordance with (*hk0*) lattice spacing of graphite (typical demonstrations are shown in the insets of Figs. 4 (b), 5(a), 6(c)). Along the (*100*) diffraction circles, some diffraction segments are also seen to be uniformly distributed (Figs.5-9(d)). Each segment, or precisely the diffraction-spot-clusters are nothing but the aggregation of a number of spots, the minimum number of which is observed to be 3 (Fig. 7(d)). The diffused nature of these diffraction spots is mostly due to the ripples present in the graphene planes (Fig. 5(b)) as has been described in the literature [15]. The angular range of each spot aggregation is observed to be constant along the different circles; their length thus increases with the radii. In the SAED pattern corresponding to Ar (Fig. 4(d)), the segments are joined; head to tail, leading to continuous circles (Fig. 4(d)) with periodic reinforcement of brightness.

These textured patterns (Figs. 4-9(d)) are very similar to those obtained for polycrystalline samples, whose cleavage planes are randomly rotated about the *c* axis, normal to the planes of the flakes so as to exhibit only (*hk0*) reflections. The positions of the segments on a particular circle are roughly staggered in radial directions with respect to those in the next circle in the diffraction pattern. The number of spots in a segment is variable and depends on the layering of the sample and is in accordance with the rotational staking faults in pyrolytic graphite. However, it is noteworthy that the AB Bernel staking seems to be absent in all the samples; rather staking of graphene layers is in accordance with the Moiré patterns with an average rotation-angle of ~20° as demonstrated in Fig. 7(d). The three spots in each segment of (*100*) diffraction circle in Fig. 7(d) can either be attributed to three graphene sheets parallel to each other and rotated through ~20° in succession about the *c* axis. As per authors' understanding, they may even be attributed to the presence of disclination (associated with the screw dislocation) in the helically wound graphitic cone. However, the TEM micrographs (Figs. 7(a-c)) associated with this kind of diffraction pattern (Fig. 7(d)) support the existence of three graphene sheets stacked one above the other.

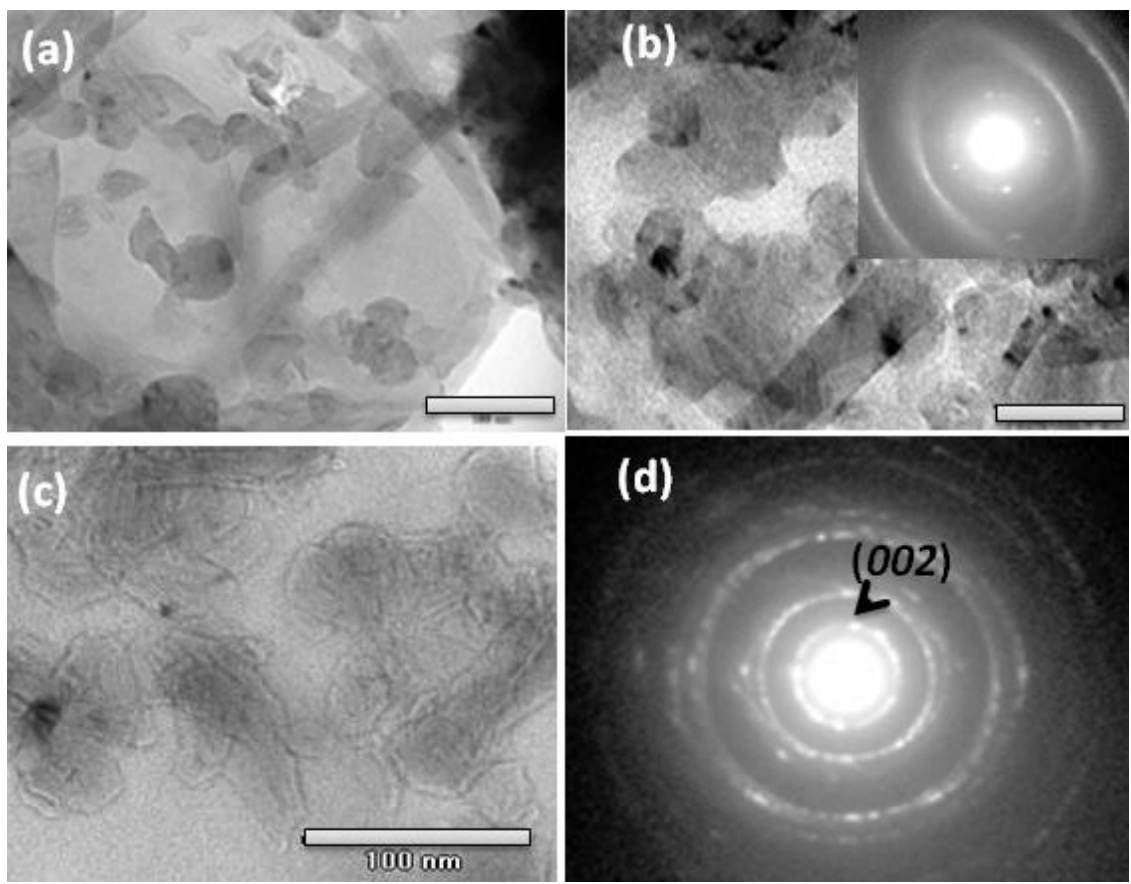

Fig.4. (a-c) Typical TEM micrographs of the sample synthesized in the ambience of 100% pure Ar and (d) the corresponding typical SAED pattern. The inset of (b) is a typical SAED pattern when the sample was tilted. The length bar equals 100 nm.

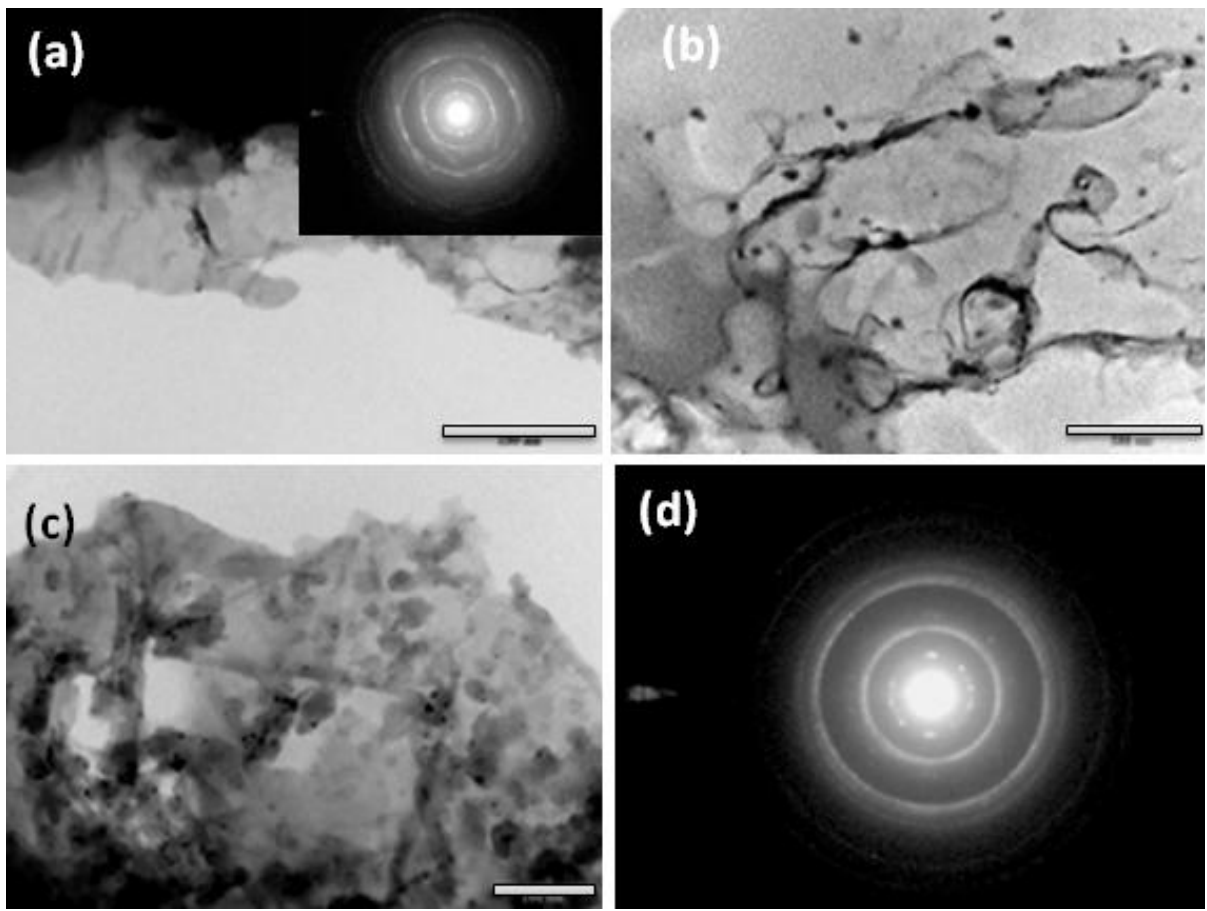

Fig.5. (a-c) Typical TEM micrographs of the sample synthesized in the ambience of 100% pure He and (d) the corresponding typical SAED pattern. The inset of (a) is a typical SAED pattern when the sample was tilted. The length bar equals 100 nm.

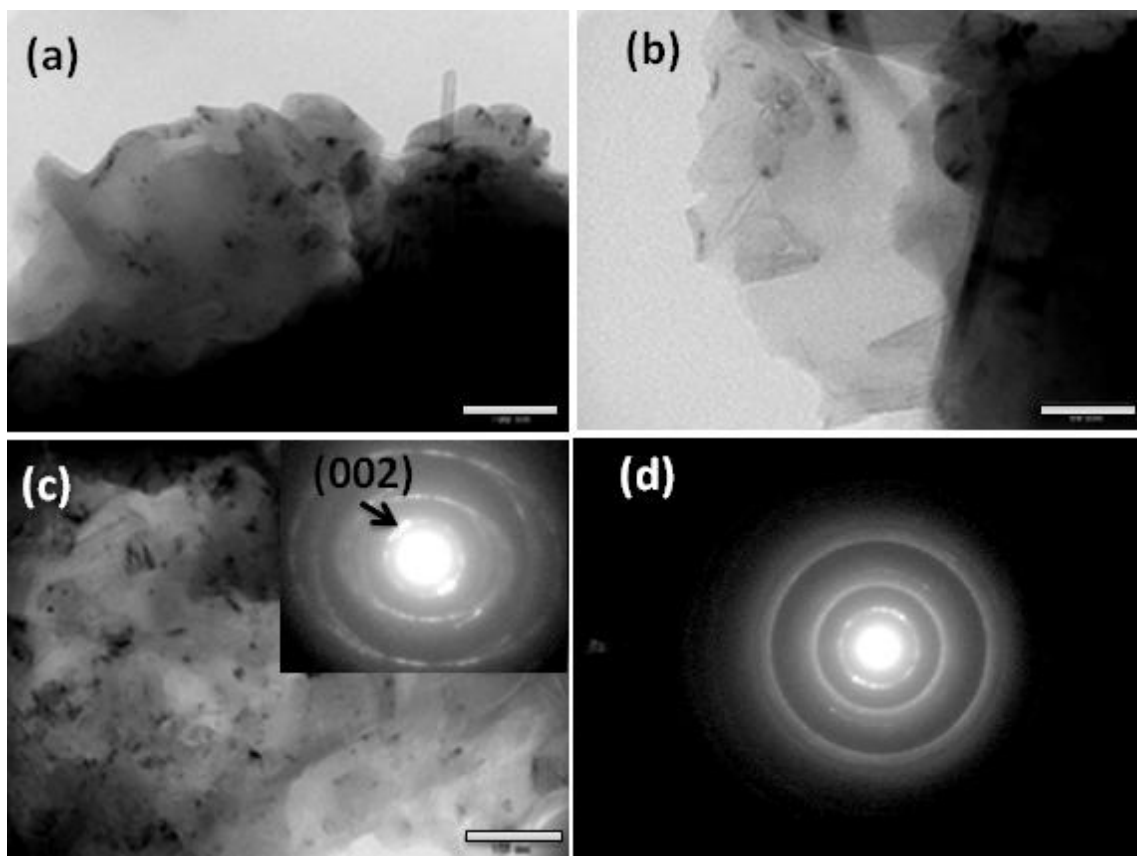

Fig.6. (a-c) Typical TEM micrographs of the sample synthesized in the ambience of 100% pure $N_2$ and (d) the corresponding typical SAED pattern. The inset of (c) is a typical SAED pattern when the sample was tilted. The length bar equals 100 nm.

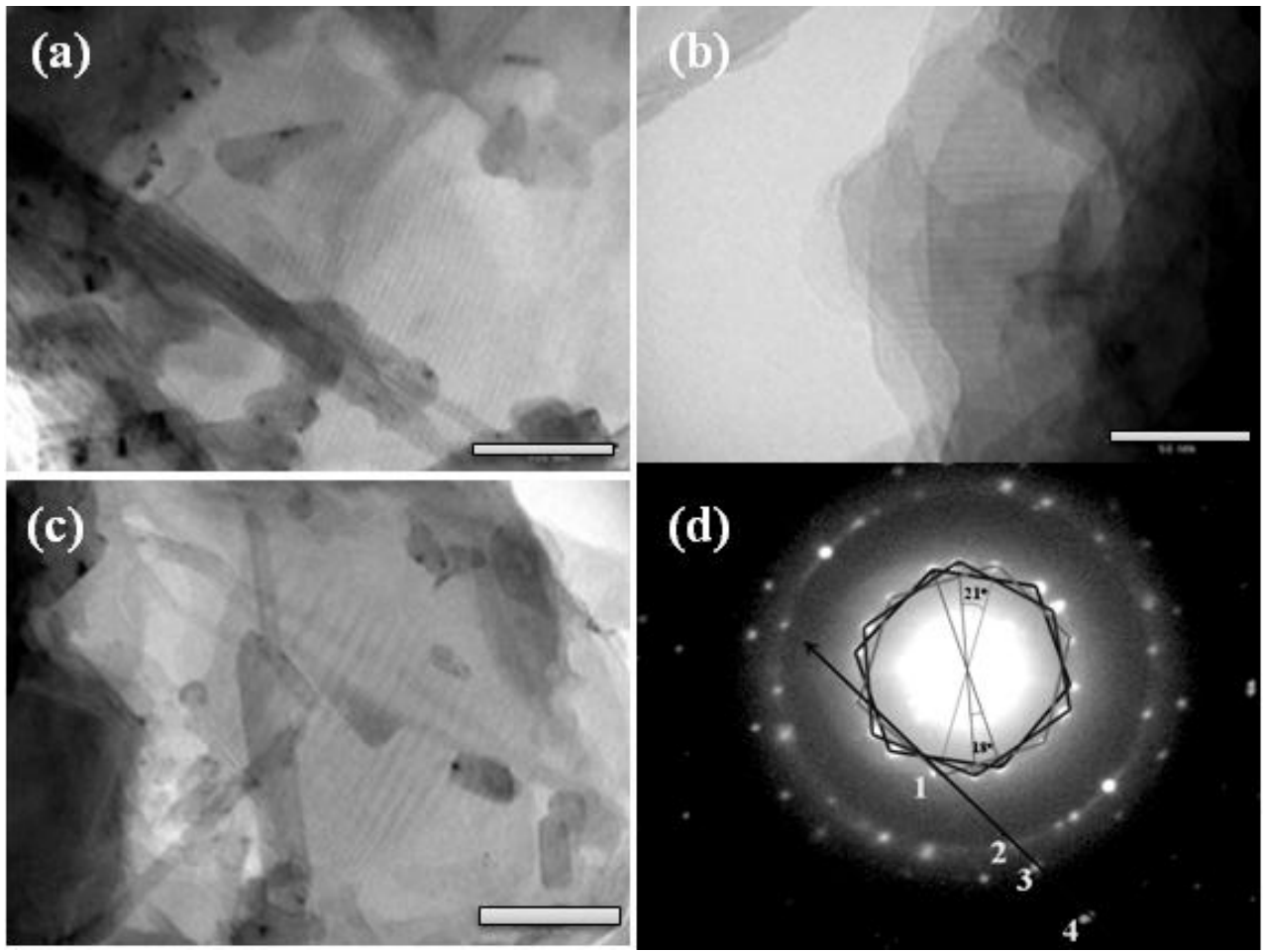

Fig.7. (a-c) Typical TEM micrographs of the sample synthesized in the ambience of 50% Ar + 50% He mixture and (d) the corresponding SAED pattern; 1,2,3 and 4 correspond to the (*100*), (*110*), (*200*) and (*210*) crystallographic planes of graphite respectively. In (d), hexagonally symmetric diffraction spots have been joined to construct three hexagons azimuthally shifted from each other. The scale bar equals 100 nm.

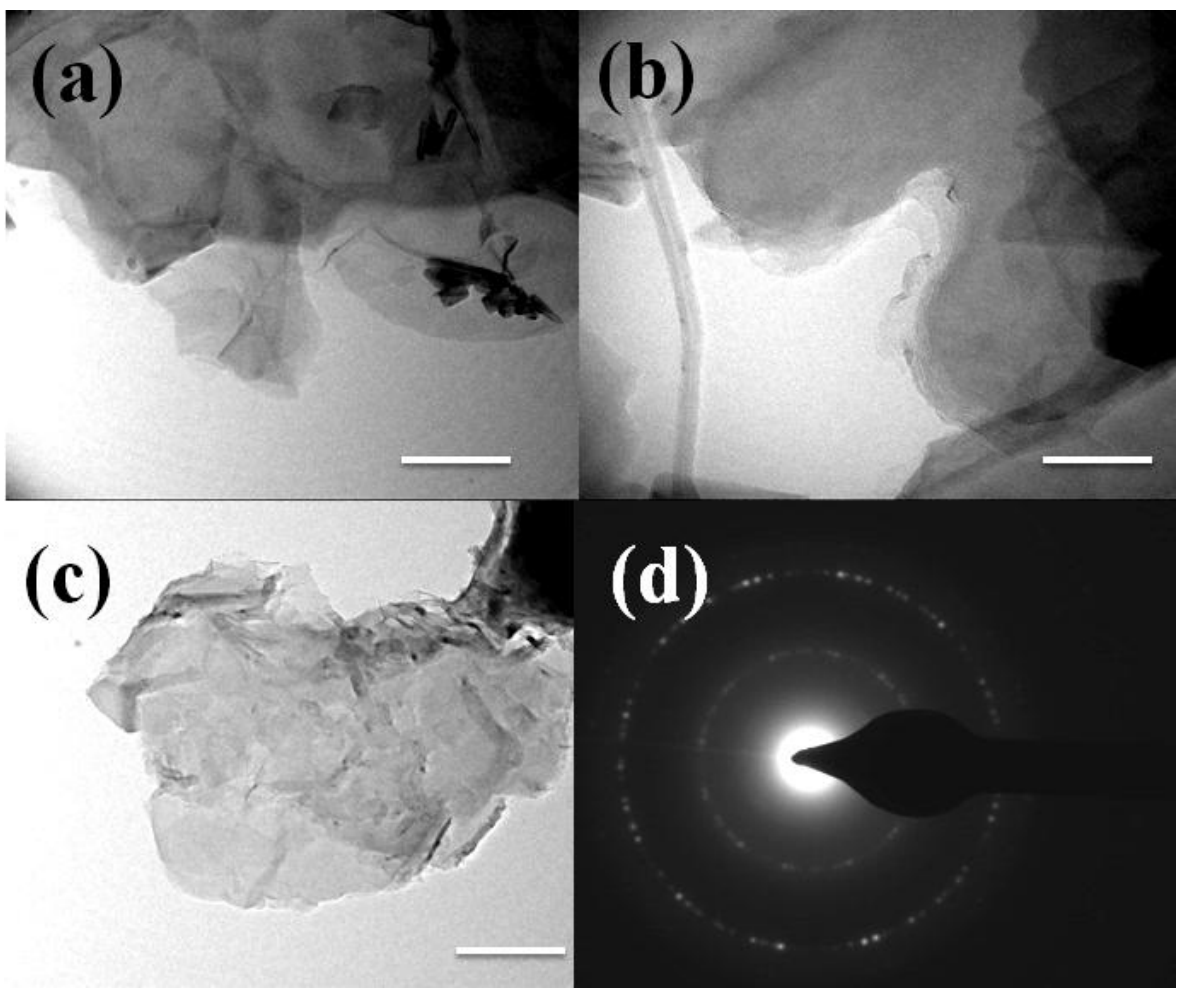

Fig.8. (a-c) Typical TEM micrographs of the sample synthesized in the ambience of 50% pure Ar + 50% pure $_{N2}$ and (d) the corresponding SAED pattern. The length bar equals 100nm.

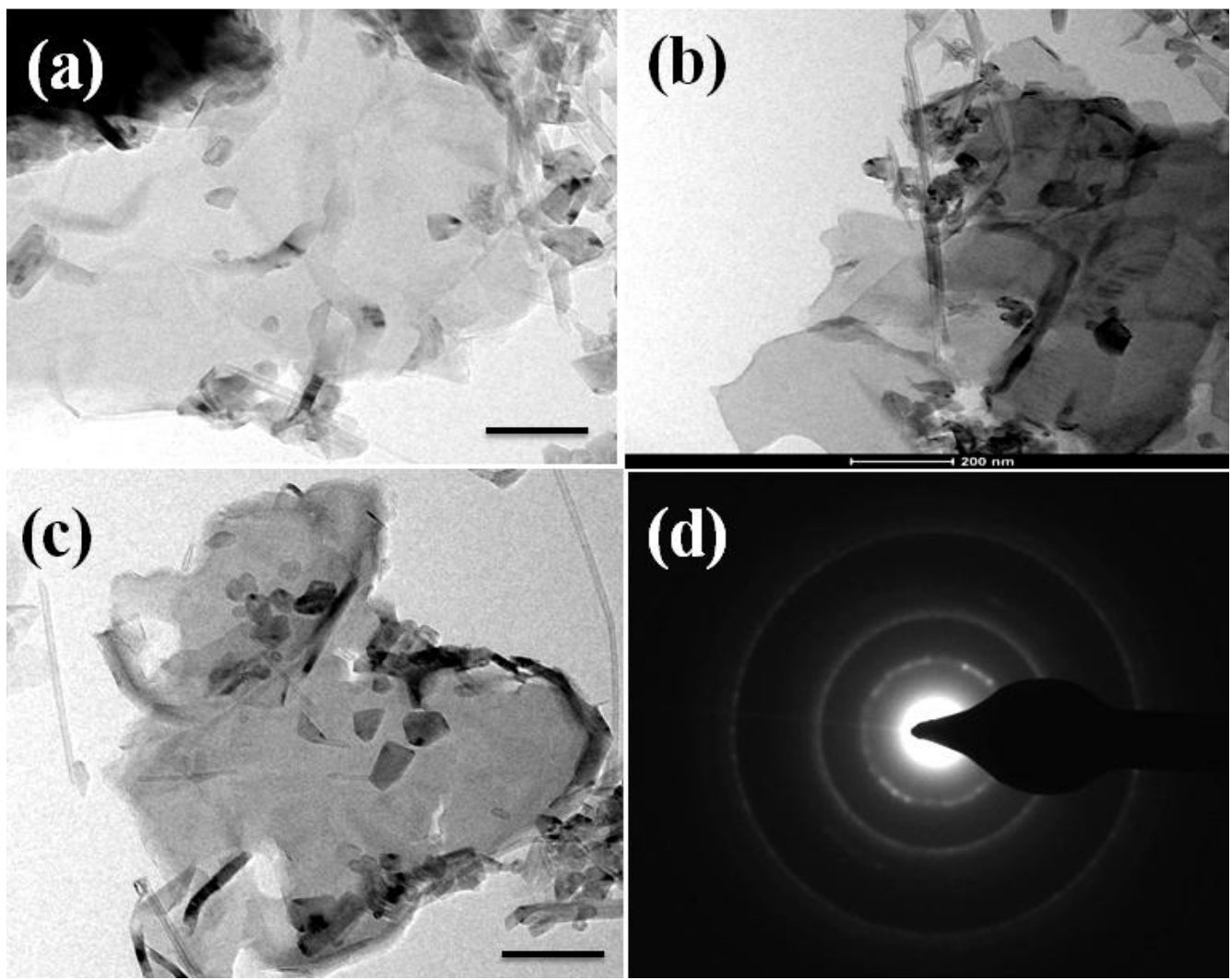

Fig.9. (a-c) Typical TEM micrographs of the sample synthesized in the ambience of 50% pure He + 50% pure $_{N2}$ and (d) the corresponding SAED pattern. The length bar equals 100nm.

The arguments put forward so far clearly indicate that the as-synthesized FLG sheets are duly delaminated and mostly multi-layered. However, effective reduction in the number of layers are observed in the samples synthesized in presence of He, the minimum number of three being noticed in case the buffer gas was a mixture of Ar and He.

*3.3 Raman spectroscopic analysis*

Raman spectroscopy has been extensively utilized for characterizing carbonaceous materials, including graphenes [16]. Excitation energy dependent dispersal of various Raman excited Stoke's lines has also been rigorously studied to understand the structure-property correlations of these materials. However, it was found that the 2D band of carbon can, in general, classify a graphene structure in a better way [16]. Moreover, it has also been found that, when reducing the excitation energy, Raman spectra of the 2D bands are more detailed and easier to identify as their internal structure spreads over a wider energy span [16]. In view of this, the excitation wavelength in the present experiment was chosen to be 633 nm to characterize the as-synthesized CDs. The Raman spectra were recorded in the range of Raman shift of 0-3000 $cm^{-1}$. However, because of the absence of any spectral feature, below 1200 $cm^{-1}$ in all the six samples, the spectra are shown in the range of 1200- 3000 $cm^{-1}$ (Fig.10).

It is noteworthy that, all the Raman spectra recorded for the as-synthesized CDs, exhibit similar nature and can hardly be distinguished from each other. An analysis of these spectra reveals that there is hardly any sample dependent dispersal of either the peak positions or the band-widths. Various parameters extracted out of these spectra are tabulated in Table 2. The physical significance of these parameters, are available in the literature [16].

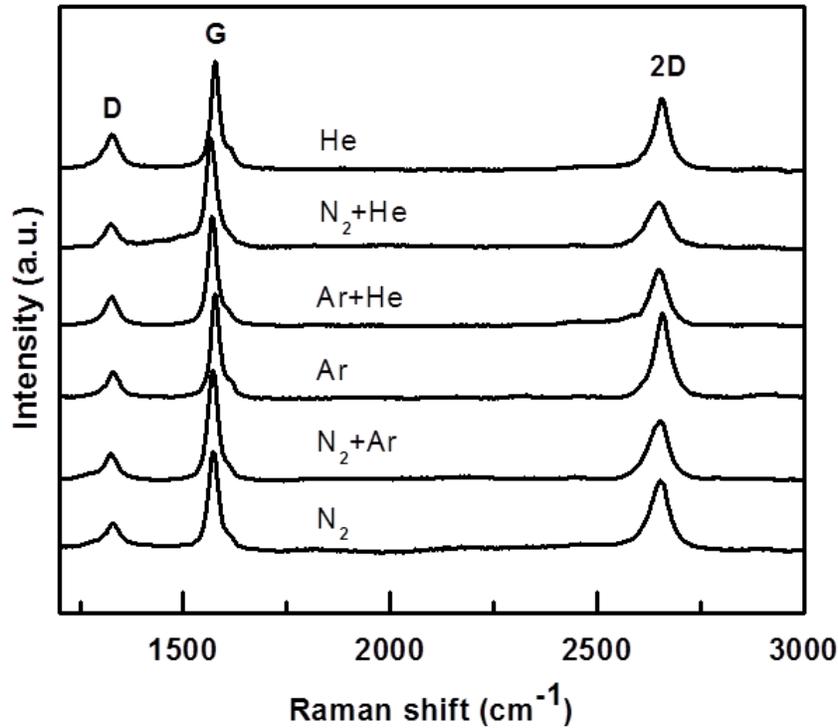

Fig.10. Comparative first and second-order Raman spectra of the as-synthesized samples.

It is to be recalled that in the Raman spectra, obtained from the samples with small crystallite size $La$ (less than 0.5 μm, i.e. smaller than the wavelength of light), the presence of D band is observed when the laser excitation energy is 2.41 eV or more. This feature is assigned to the breathing of the carbon hexagons that become Raman active at the borders of the crystallite areas owing to the loss of translational symmetry [17, 18]. This peak may even arise due to presence of defects in an otherwise perfect graphite crystal with large dimension. Therefore, the presence of D band in all the spectra reveals the existence of both the aforesaid possibilities in the as-synthesized graphitized samples.

Table 2. Parameters extracted from the Raman spectra of the as-synthesized samples.

| Buffer gas | D peak position (cm$^{-1}$) | G peak position (cm$^{-1}$) | 2D band position (cm$^{-1}$) | $I_G/I_D$ ratio (Integrated area ratio) | $I_{2D}/I_G$ ratio (Integrated area ratio) | G band line width (cm$^{-1}$) | 2D band line width (cm$^{-1}$) |
|---|---|---|---|---|---|---|---|
| Ar, He, N$_2$, 50%Ar+50%He, 50%Ar+50% N$_2$, 50% N$_2$+50%He | 1326.9±2.6 | 1572.2±3.9 | 2653±3.2 | 3.4±0.4 | 1.23±0.2 | 25±1.6 | 50.4±5.7 |

Moreover, all the six D bands observed herein could be fitted to a single Lorentzian and showed the complete absence of the doublet-feature. This observation is indeed significant and clearly indicates that all the six samples are completely free from the coupling in the two adjacent graphene layers stacked in a highly ordered ……ABABAB…. format [19].

Moreover, in the present experiment $I_G/I_D$ ratio is seen to decrease to a value of about 3.4 from a value of 10 in our previous report [10]. This may either be attributed to more exposed edge states and wrinkles, or reduction in the crystallite dimension of the as-synthesized graphitized samples; which in turn, might be due to the inherent arc fluctuation and temperature gradient present in the reactor. The alteration in the *J*×*B* force [10] in the present experiment might also be one of the factors to be considered.

The disorder-induced band centered at *ca.* 1620 cm$^{-1}$ is usually observed in the Raman spectra of disordered graphitic materials [20]. The contribution of this peak (as was demonstrated in Ref. [12]) for all the six samples has been estimated to be lesser than 5% in the G band with negligible standard deviation. This indicates that the intrinsic structural defects of the as-synthesized FLG are minimal. This estimated value of 5% also rules out the possibility of considering the carbon nanocrystalline particles to be the major products in the present experiment. In view of the above discussion, the in-plane dimension of the as-synthesized graphenes appears to be lesser than what was observed in our previous report and the TEM micrographs (Figs.4-9, [10]) strongly support this conclusion. Using the value of the $I_G/I_D$ ratio, the crystallite size $L_a$ of the as-synthesized FLG sheets is estimated to be ~136 nm using the following equation reported in the literature [21].

$$L_a(nm) = (2.4 \times 10^{-10})\lambda_{laser}^4 \left(\frac{I_D}{I_G}\right)^{-1}$$

Rapid quenching, due to the existing steep temperature gradient in the reactor, might have arrested the growth of the as-synthesized FLG much earlier than what was observed in our previous case. There might be effects of the alteration in the ***J***×***B*** force even. Moreover, it may also be noted that the position of the G band is seen to be red-shifted by ~10 cm$^{-1}$ from the one reported for HOPG [16]. This red-shift might be due to the domination of extra strain induced as a result of their misorientaion over the strain relaxation along the *c* direction (section 3.2) in the adjacent layers inside the as-synthesized FLG [22]. It is finally inferred from the first order Raman spectra that neither the degree of graphitization nor the crystallite dimension of the as-synthesized carbonaceous materials have been remarkably influenced by the buffer gases. Apart from these, the first order Raman spectra did not show any other particularity.

The second order 2D feature is not a disorder-induced feature but it can nevertheless be used to probe changes in the electronic and vibrational structure related to disorder. The two-dimensional versus three-dimensional stacking order of graphene layers is one example where the 2D band provides important information. Moreover, highly crystalline three-dimensional graphite is reported to exhibit two 2D peaks [16].

A single 2D band is observed in all our samples, however with a global red-shift at *ca.* 1653 cm$^{-1}$. This global red-shift may be attributed to stress relaxation due to mechanical decoupling along the *c* axis of individual layer of the as-synthesized FLG and is consistent with the SAED features observed in case of our samples (insets of Figs. 4(b), 5(a), 6(c)). The absence of the two-peak nature of the 2D band also indicates that the as-synthesized samples are mostly free from three-dimensional crystalline graphite. It should also be noted that the 2D peak position is in accordance to the value reported in the literature [23] for the same exciting laser wavelength used in the present study and clearly indicates the presence of misoriented graphene layers inside the as-synthesized samples.

In some literature [24], the $I_{2D}/I_G$ ratio has been shown to be the signature, which can be utilized to quantify the number of layers in FLG. In a single layer graphene, 2D peak intensity has been observed to be more than the G band [24]. However, graphite polyhedral crystals also give rise to the same feature in the Raman spectra [19]. Apart from this, the degree of suspension also affects this ratio; higher the degree of suspension, higher is the ratio [22]. Therefore, in analyzing the powdered sample like ours, the $I_{2D}/I_G$ ratio may not be a parameter to give much importance to, and unambiguous qualification of the as-synthesized samples may not be possible using this ratio. However, the value of this ratio, as furnished in Table 2 is well in accordance with the value reported for the FLG [22]. Moreover, the standard deviation in the 2D peak position, as furnished in Table 2, may be a signature of the unevenness of the sample thickness or varying number of layers.

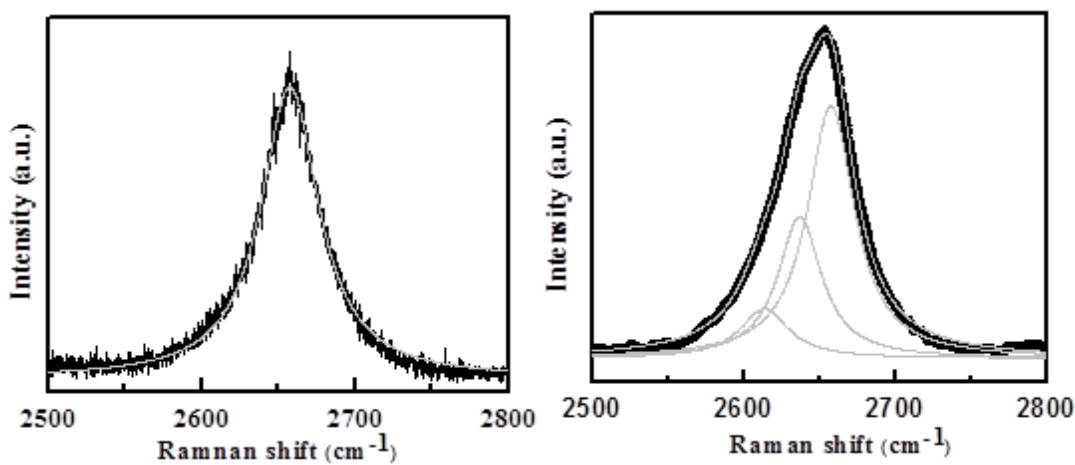

Fig.11. Typical 2D bands observed in case of all the six samples on subsequent position dependent recording of the Raman spectra and their fittings by Lorentzian line-shapes. The dark lines are the actual spectra; whereas, the gray lines are the fitted ones.

While de-convoluting the 2D band into Lorentzian line-shapes, it is seen to be de-convoluted either in a single or triple Lorentzians as furnished in Fig. 11. However, none of the Raman spectra recorded for the as-synthesized samples showed the existence of two peaks in the 2D band clearly indicating that most of the constituents of the samples are not three-dimensionally stacked. This finding also strengthens the belief that all the as-synthesized samples are mostly FLG dominated as is supported by the TEM micrographs (Figs. 4-9). The number of Lorentzian line-shapes was not found to be an unambiguous character of a particular sample; rather they were observed to vary in case of all the samples quite randomly in varying the location of the laser spot. The fitting also indicates [16, 22, 24] that all the samples are mostly dominated by FLG structures and there is a distinct possibility of finding graphene sheets in the samples, which might have been exposed to the incident laser out of the delaminated structures, a characteristic feature of this kind of magnetic field modulated carbon arc deposit.

*3.4 XRD analysis*

A typical XRD pattern corresponding to all the mechanically homogenized CDs is shown in Fig. 12 and it is similar to that observed for well graphitized carbon samples [25]. No signature of the inclusion of N atom in the as-synthesized graphitic structures, produced in the $N_2$ ambience was observed.

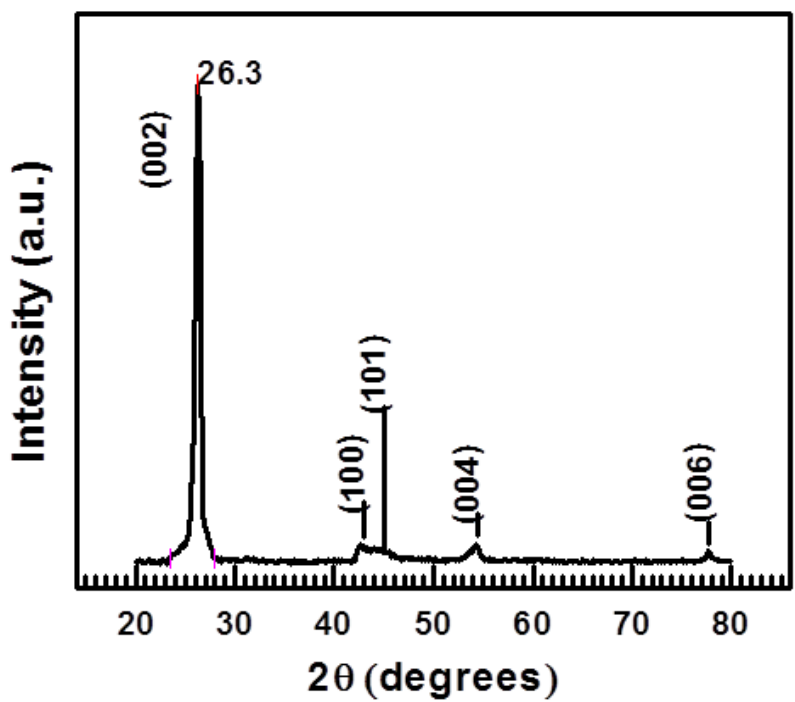

Fig.12. Typical XRD pattern associated with all the as-synthesized samples.

Although, the XRD profiles are not useful to differentiate the constructional details between the graphene and the graphitized structures; still, they can help to determine the nature of crystallinity and the purity of the samples. The typical XRD patterns obtained for all the as-synthesized samples (Fig.12) indicate a good degree of purity and verify the crystallinity of the samples. The XRD peak at $2\theta = 26.4°$ is a characteristic of graphite [25]; a red-shift of the position of this peak, as is found in Fig.12 is attributed to enhanced inter-layer separation and indicates de-lamination of the graphitic layers. This result is consistent with the elliptical diffraction patterns (insets of Figs. 4(b), 5(a), 6(c)) and the arguments resulted thereof previously [10]. No buffer gas dependent dispersal of the peak position, intensity ratio and the line-width was observed indicating that all the as-synthesized samples are of comparable qualities in terms of both inter-planer separations and crystallinity.

*3.5 Growth mechanism*

In view of the above findings, the following growth mechanism of the as-synthesized FLG appears to be quite reasonable. First, due to the arc rotation, the temperature distribution in the vicinity of the cathode surface becomes homogeneous. The cathode surface, being below the graphite sublimation temperature in the present case, provides a favorable zone for the nucleation of thermodynamically stable graphitic structures. The directed flux of the carbon atoms from the anode to the cathode, which are responsible for the formation of CNT [12, 13] becomes diffusive in the radial direction on account of the imposed $\boldsymbol{J}\times\boldsymbol{B}$ force [10] and the stacking of carbon atoms along the cathode surface becomes more favorable. This produces a situation which guides the carbon precursors to nucleate first in a single layer of graphene growing further in the $\boldsymbol{J}\times\boldsymbol{B}$ direction. With time, this seed graphene layer further grows along the direction of the arc axis on account of two competitive processes *viz.* loss of carbon atoms from the as-grown graphene surface by the process of sublimation and absorption of the adatoms from the impinging carbon flux. The later process is responsible for the stacking of another layer to the seed graphene layer thereby making it a bi-layer graphene. This process continues further; however, newly generated graphene layers stack to the seed layer at azimuthally displaced positions due to the arc rotation. In this way, rotational stacking faults become inherent parts of the as-grown FLG. This is evident from the SAED-pattern-analysis as discussed before (section 3.2). The growth finally stops as the temperature of such FLG decreases below its optimum growth-temperature on account of the thermal conduction through the buffer gas medium. The buffer gas with higher thermal conductivity will cease the growth earlier than a medium with low thermal conductivity value. As a result, the number of layers of the as-grown FLG is expected to be less in a medium with high thermal conductivity. Possibly, this is why the number of layers of such FLG was found to be minimum in presence of He (section 3.2), which has a much higher value of thermal conductivity as compared to both $N_2$ and Ar. However, the impinging carbon flux, as pointed out before, is also an important factor to be analyzed critically before the exact structures of the as-grown FLG can be prescribed unambiguously. An optimization of the impinging carbon flux, ***J*×*B*** force and the thermal conductivity of the buffer gas can thus produce the most favorable situation for the formation of FLG with rotational stacking faults, pretty high crystallite dimension and minimum number of layers.

## 5. Conclusions

The CDs, formed in presence of six different non-reactive buffer gases (100% Ar, 100% He, 100% $N_2$, 50% Ar+50% He, 50% Ar+50% $N_2$, 50% $N_2$+50% He), in an external magnetic field modulated DC carbon arc were investigated in detail without any post-synthesis-treatment. The major findings of this study are as follows.

The rates of anode-erosion and CD-formation are greatly affected by the thermo-physical properties of the buffer gases. It was found that these rates are lowest in case of Ar, while the highest values correspond either to He or $N_2$, the intermediate values being attributed to the mixed buffer gases. In view of this, enthalpy of the arc appears to be the most dominant factor, which decides the carbon flux responsible for the formation of nanocystalline structures on the cathode surface.

The diameter of the CD, which is a parameter to externally monitor the formation of FLG inside the CD in this kind of arc [10], is unaffected by the nature of the buffer gases. Rather, the ***J*×*B*** force is the only parameter, which can control the CD-diameter for a used set of electrodes.

The soft cores of the as-synthesized CDs are found to contain mostly FLG sheets along with traces of closed graphitic structures *viz.* CNTs and carbon particles as is evident from the TEM micrographs. The formation the closed graphitic structures was observed when pure buffer gases were used. However, on using mixed buffer gases, FLG sheets were found to dominate in the as-synthesized products. An analysis of the SAED patterns corresponding to these micrographs clearly indicate the presence of multiple delaminated graphene layers with hexagonal symmetry stacked one above the other; however, with rotational stacking faults. This was further reconfirmed with the help of 2D band analysis of the Raman spectra. The layers of such FLG sheets were found to be mostly ABC stacked and were found to decrease in number on using mixed buffer gases containing He.

The in-plane dimension of the as-synthesized FLG, as revealed through both TEM and Raman analysis, is unaffected by the nature of the buffer gases; rather, $\boldsymbol{J}\times\boldsymbol{B}$ force, arc instabilities and the temperature gradients appear to be the factors to be given proper attention to.

The FLG sheets, synthesized by this method, bear a good degree of graphitization and less intrinsic defect-states as was confirmed both by Raman spectroscopy and XRD analysis.

It is therefore concluded that a duly optimized external magnetic-field-modulated DC carbon arc can be envisaged as an easy route to synthesize FLG with rotational stacking faults with a g/min production rate in an environment-friendly manner, without liberation of any kind of toxic or hazardous material. Such FLG sheets can find potential applications in a large number of composite materials and science, as well as in batteries.

**Acknowledgements**

The financial support from the Department of Atomic Energy, Govt. of India previously obtained for fabricating the arc reactor is gratefully acknowledged. SK acknowledges Indrani Banerjee for supplying few important articles and some non-academic support from the management of the

Birla Institute of Technology (Mesra), Deoghar Campus, to carry out this research work. SVB acknowledges CSIR, India, for granting her the Emeritus Scientist project.